\documentclass[12pt]{iopart}

\expandafter\let\csname equation*\endcsname\relax
\expandafter\let\csname endequation*\endcsname\relax 

\usepackage{graphicx} 
\usepackage{lmodern} 
\usepackage{textcomp} 
\usepackage[T1]{fontenc} 
\usepackage{gensymb} 

\usepackage{graphicx,amsfonts,amsmath,amssymb,mathrsfs}
\usepackage{dcolumn}
\usepackage{bm}
\usepackage[usenames]{color}

\usepackage{epstopdf}

\begin{document}

\newcommand{\figref}[1]{Fig.~\ref{#1}}

\newcommand*{\Real}{\Re{\rm e}}
\newcommand*{\Imag}{\Im{\rm m}}
\newcommand*{\ket}[1]{|{#1}\rangle}
\newcommand*{\bra}[1]{\langle{#1}|}
\newcommand*{\Bra}[1]{\left<#1\right|}
\newcommand*{\Ket}[1]{\left|#1\right>}
\newcommand*{\mean}[1]{\mathinner{\langle{#1}\rangle}}
\newcommand*{\braket}[2]{\mathinner{\langle{#1}|{#2}\rangle}}
\newcommand*{\ketbra}[2]{\mathinner{|{#1}\rangle\langle{#2}|}}
\newcommand*{\xbraket}[3]{\mathinner{\langle{#1}|{#2}|{#3}\rangle}}
\newcommand*{\twobytwo}[4]{\begin{pmatrix} #1 & #2 \\ #3 & #4\end{pmatrix}}
\newcommand*{\vectwo}[2]{\begin{pmatrix} #1 \\ #2 \end{pmatrix}}
\def\be{\begin{equation}}
\def\ee{\end{equation}}

\newcommand*{\Fst}{\Psi_\alpha} 
\newcommand*{\Fmd}{\varphi_\alpha} 
\newcommand*{\Fqe}{\epsilon_\alpha} 
\newcommand*{\Fnoise}{X_{\alpha,\beta,k}} 
\newcommand*{\Fgap}{\mathcal {E}} 
\newcommand*{\varphiS}{\tilde{\varphi}_S} 
\newcommand*{\Dabk}{\Delta_{\alpha\beta,k}} 
\newcommand*{\Aabl}{A_{\alpha\beta}^\nu} 
\newcommand*{\aabk}{a_{\alpha\beta,k}} 
\newcommand*{\nth}{n_{\rm th}} 
\newcommand*{\pup}{p_\uparrow} 
\newcommand*{\pdn}{p_\downarrow} 


\newcommand{\gap}{\Omega_R}
\newcommand{\average}[1]{\langle #1 \rangle}
\newcommand{\CF}{ G(\nu,t)}
\newcommand{\st}{{\rm st}}
\newcommand*{\CFtext}{characteristic function}
\newcommand{\matrixel}[3]{{\mathinner{\langle{#1}| {#2} | {#3}\rangle}} }

\newcommand*{\rhoI}{\rho^{(I)}}
\newcommand*{\Hlambda}{H_{\frac{\nu}{2}}}
\newcommand*{\Hmlambda}{H_{-\frac{\nu}{2}}}
\newcommand*{\Vlambda}{V_{\frac{\nu}{2}}}
\newcommand*{\Vmlambda}{V_{-\frac{\nu}{2}}}

\newcommand*{\rhoT}{\rho_T}
\newcommand*{\rhoS}{\rho}


\title{Heat-exchange statistics in driven open quantum systems}
\author{S Gasparinetti$^{1}$, P Solinas$^{2}$, A~Braggio$^{3}$ and M.~Sassetti$^{2,3}$}
\address{
$^1$ Low Temperature Laboratory (OVLL), Aalto University, P.O. Box
15100, FI-00076 Aalto, Finland}
\address{$^2$ SPIN-CNR, Via Dodecaneso 33, I-16146 Genova, Italy}
\address{$^3$   Dipartimento di Fisica, Universit\'a di Genova, Via Dodecaneso 33, 16146 Genova, Italy}
\ead{simone.gasparinetti@aalto.fi}

\begin{abstract}
As the dimensions of physical systems approach the nanoscale, the laws of
thermodynamics must be reconsidered due to the increased importance of
fluctuations and quantum effects.
While the statistical mechanics of small classical systems is relatively well
understood, the quantum case still poses challenges.
Here we set up a formalism that allows us to calculate the full probability
distribution of energy exchanges between a periodically driven quantum system
and a thermalized heat reservoir.
The formalism combines Floquet theory with a generalized master equation
approach.
For a driven two-level system and in the long-time limit, we obtain a universal
expression for the distribution, providing clear physical insight into the
exchanged energy quanta.
We illustrate our approach in two analytically solvable cases and discuss the
differences in the corresponding distributions.
Our predictions could be directly tested in a variety of systems, including
optical cavities and solid-state devices.
\end{abstract}

\date{\today}

\maketitle

\section{Introduction}

In small systems, fluctuations of thermodynamic quantities become important.
The study of fluctuations in small systems led to the development of stochastic
thermodynamics \cite{Seifert2005, Seifert2012}, whose predictions, such as
the Jarzinski equality \cite{Jarzynski1997} and the Crooks fluctuation relations
\cite{Crooks1999}, have been experimentally verified in mechanical
\cite{Wang2002}, biological \cite{Liphardt2002}, molecular
\cite{Collin2005,Alemany2012}, optical \cite{Schuler2005}, and electronic
systems \cite{Saira2012a}. Thermal fluctuations can also be exploited to extract
work by using nonequilibrium feedback, allowing for physical realizations of the celebrated
Maxwell's demon \cite{Szilard1929,Sagawa2010,Toyabe2010,Koski2013,Koski2014}.

In all the mentioned experiments, the underlying physics is essentially classical.
Down to atomic size and low temperatures, however, quantum mechanics is expected
to come into play.
Thermodynamics and statistical mechanics of quantum systems,
as well as different types of quantum fluctuation relations and their observability in different settings,
have been the subject of much attention lately \cite{Campisi2011, Huber2008, Campisi2009, Heyl2012, Dorner2013, Mazzola2013,Campisi2013}.
However, while experimental demonstrations are still lacking, the very definition of basic concepts, such as, work, is still under debate
\cite{Allahverdyan2005, Talkner2007, Campisi2011, Solinas2013}.
This difficulty is related to the measurement process: on the one hand, work
is not by itself an ``observable'' in quantum-mechanical sense; on the other
hand, its definition in terms of a double projective measurement, which
prevails in literature \cite{Campisi2011}, involves coherence loss between
different-energy eigenstates after the first measurement. An interferometric measurement of the work distribution using an an ancilla was considered
in \cite{Dorner2013, Mazzola2013}. It has also been proposed to measure the
environment, rather than the system, to keep track of the energy exchanges between
the two \cite{Pekola2013,Hekking2013,Gasparinetti2014}.

Here, we provide a general framework to
understand the statistics of energy exchanges between a periodically driven quantum system and
a heat bath.
The problem of dissipation in driven systems has a long story \cite{Weiss1993,Grifoni1996,Grifoni1998,Breuer2007}.
However, the statistics of the dissipated heat was considered only recently and either in very specific cases
and/or in the weak-drive limit \cite{Solinas2013, Hekking2013, Silaev2014}.
We set the stage by deriving from microscopic principles a generalized master
equation describing the dynamics of the energy exchanges.
We focus on a driven two-level system; under a limited set of assumptions, we
provide analytical expressions for the mean heat power and for all the cumulants
of the energy distribution after a given time.
In the long-time limit, we also write down the full probability distribution
in a general form. By highlighting the contribution of each individual energy
exchange, the full distribution provides us with a deeper insight into the
thermalization process.

Within the stated assumptions, our results apply to any type of drive and
coupling mechanism.
Among different regimes, that of strong driving is particularly interesting, as
it cannot be addressed with pertubative techniques -- the system and the drive
``hybridize'' and thermalization takes place by exchanging photons at
dressed-state energies.
We illustrate these concepts by discussing two analytically solvable cases, in
which the same driven quantum system is coupled to the environment in two
different ways. As we show, different coupling mechanisms can result in very
different energy-exchange processes.
Our formalism can be straightforwardly generalized to consider multi-level systems and multiple heat baths; as such, it could be
used to study the performance of quantum heat pumps and engines
\cite{Kosloff2014,Alicki1979,Kosloff1984,Geva1992,Scully2002,Scully2003,Segal2006,Erez2008, Abah2012, LevyPRL2012, LevyPRE2012,Gelbwaser-Klimovsky2013,Rossnagel2014}.

\section{Theoretical framework} 
We consider a generic quantum system subject to a periodic drive and weakly coupled to its environment.
The environment is a macroscopic object, which we assume to be thermalized at all times.
Our quantity of interest is the probability density function (PDF) $P(Q,t)$
for the environment to exchange the amount of energy $Q$ between times $0$ and
$t$.
The moments of $Q$ are conveniently expressed in terms of the characteristic
function (CF) $\CF=\int_{-\infty}^{\infty} dQ P(Q,t) e^{i Q
\nu}\ $, as follows: $\average{Q^n(t)} = (-i)^n \left.\frac{d^n
\CF}{d\nu^n}\right|_{\nu=0}$.

We write the total Hamiltonian as $H_T(t)=H(t) + H_R + V$,
where $H(t)$ and $H_R$ are the system and reservoir Hamiltonians, respectively,
and $V=S \otimes B$ is a coupling term consisting of operators $S$ and $B$
acting in the Hilbert space of system and reservoir, respectively.
In the following we will take $\hbar=1$ and denote with $\rhoT$, $\rho$ and $\rho_R$ the total (system $+$ environment),
system and environment density operators, respectively.

The PDF for the heat amount $Q$ to be transferred to the
reservoir between times $t_0=0$ and $t$ is given by \cite{Talkner2009,Esposito2009}
\be 
P(Q,t)=\sum_{e_1,e_2} \delta(e_2-e_1-Q) p[e_2;e_1] p[e_1] 
\label{eq:p(q)} \ , 
\ee 
where $p[e_2;e_1]$ is the conditional probability that a
measurement of $H_R$ gives $e_2$ at time $t$ when it gave $e_1$ at time $t_0$
and $p[e_1]$ is the probability to measure $e_1$ at time $t_0$.
Introducing the projector $P_{e_j}$ on the $j$-th state of the reservoir of
energy $e_j$ and using the property $P_{e_j}^2=P_{e_j}$, we have
\be
\begin{split}
p[e_2,e_1] p[e_1] &= \Tr [P_{e_2} U(t) P_{e_1} \rhoT(0) P_{e_1} U^\dagger(t) P_{e_2}]   \\
 		             &= \Tr [U^\dagger(t) P_{e_2} U(t) P_{e_1} \rhoT(0) P_{e_1}] .
\end{split}
\ee
where $U(t)$ is the evolution operator generated by $H_T(t)$.
The CF is given by \cite{Esposito2009}
\be
\CF \equiv \int_{-\infty}^{\infty} dQ P(Q, t) e^{i Q \nu} = \sum_{e_1,e_2} p[e_2;e_1] p[e_1]e^{i \nu (e_2-e_1)} \ . \label{eq:G_def}
\ee

We assume the initial total density matrix $\rhoT(0)= \rho(0) \otimes \rho_R(0)$ to be factorized into the system density matrix $\rho(0)$ and the thermalized environment density matrix $\rho_R(0)= e^{-\beta H_R}/Z_R$, where $Z_R$ is the partition function of the environment.
In particular, this assumption implies that all projectors $P_{e_j}$ commute with $\rho(0)$, so that the initial measurement of the observable of interest (here, $H_R$) does not change the subsequent dynamics \cite{Solinas2013, Esposito2009}.

After noticing that $\sum_{e_j} P_{e_j} e^{\pm i \nu e_j}= e^{\pm i \nu H_R}$,
we write \eqref{eq:G_def} as
\be
\CF = \Tr [U^\dagger(t) e^{i \nu H_R} U(t) e^{-i \nu H_R} \rhoT(0)] 
\ee
and finally as
\be
\CF= \Tr [\rhoT^\nu(t)]  \ ,
\ee
where
\be
\rhoT^\nu(t)=U_{\nu/2}(t) \rhoT(0) U^\dagger_{-\nu/2}(t)
\label{eq:rho_T_def}
\ee 
and 
\be
U_\nu(t)=e^{i \nu H_R} U(t) e^{-i \nu H_R}  \ 
\ee
satisfies the equation of motion $i dU_\nu(t)/dt =
H_\nu(t) U_\nu(t)$, with $H_\nu (t)= e^{i\nu
H_R}H_T(t)e^{-i\nu H_R}$. 
The equation for $\rhoT^\nu(t)$ can be compared to similar expressions obtained from the full-counting statistics theory of charge transport \cite{Kindermann2003}, with $\nu$ playing the role of the counting field.
In general, the evolution of $\rhoT^\nu$ is not unitary for $\nu \neq 0$.

For a periodic drive, $H(t+\tau)=H(t)$, where $\tau$ is the drive period and
$\Omega=2\pi/\tau$ the corresponding angular frequency.
We assume that the unitary dynamics generated by $H$ is known; this knowledge is
captured in our formalism by the use of Floquet states and quasienergies
\cite{Grifoni1998}.
Floquet states are particular solutions of the Schr\"odinger equation of the
form $
\ket{\Fst(t)}=e^{-i\Fqe t}\ket{\phi_\alpha(t)}
$ where the Floquet mode $\ket{\phi_\alpha(t)}$ satisfies
$\ket{\phi_\alpha(t+\tau)}=\ket{\phi_\alpha(t)}$ and $\Fqe$ is its corresponding
quasienergy.


The derivation of the GQME for the generalized density matrix of the system
$\rho^\nu=\Tr_R[\rho_T^\nu]$ is done in complete analogy to the
standard master equation \cite{Breuer2007} and it is presented in \ref{app:MEQ_derivation}.
We assume weak coupling between system and environment and fast autocorrelation
time of the environment (Born-Markov approximation).
We model the environment as a collection of harmonic oscillators with bosonic
occupation $n_B(\omega)$ and Ohmic spectral density $J(\omega) = \eta \omega$
with $\eta$ a dimensionless coupling coefficient.
The so-obtained GQME can be simplified in different manners, depending on the
time scales of the problem.
In the fast-driving regime \cite{Grifoni1998, Grifoni1995, Russomanno2011,
Gasparinetti2013}, one can safely neglect oscillating terms of the form $e^{i k
\Omega t}$, with $k\neq 0$, and, in most instances, also of the form $e^{ i
(\epsilon_\alpha - \epsilon_\beta) t}$, with $\alpha\neq\beta$.
This amounts to a full secular approximation \cite{Gasparinetti2013}, which
admits an analytical solution.

\section{General results}
We write the resulting GQME in the Schr\"odinger
picture and in the Floquet basis.
Under the stated assumptions, the GQME is Markovian and time-independent.
The populations $\rho_{\alpha\alpha}^\nu=
\xbraket{\Psi_\alpha}{\rho^\nu}{\Psi_\alpha}$ are decoupled from the
coherences and satisfy a vector equation of the form $\dot{\vec{\rho^\nu}} =
\mathcal{A} \cdot \vec{\rho^\nu}$. For a two-level system, $\vec{\rho^\nu} =
\{\rho_{11}^\nu, \rho_{22}^\nu\}$
and
\be \mathcal{A} =
\begin{pmatrix}
A_{11}^\nu-A_{11}-A_{21} & A_{12}^{\nu} \\
A_{21}^{\nu} & A_{22}^\nu-A_{22}-A_{12}
\end{pmatrix} \ .
\label{eq:gqme_fsa}
\ee
A derivation of \eqref{eq:gqme_fsa} can be found in \ref{app:GQME_periodic}.
To define the coefficients in \eqref{eq:gqme_fsa}, let us first write the matrix
elements of the coupling operator $S$ in the Floquet basis \cite{Grifoni1998} as
$\xbraket{\Psi_\alpha(t)}{S}{\Psi_\beta(t)} = \sum_{k}
S_{\alpha\beta,k}e^{ i \Dabk t}$ where $S_{\alpha\beta,k}=\frac1{\tau}
\int_0^\tau dt \xbraket{\phi_\alpha(t)}{S}{\phi_\beta(t)} e^{- i k \Omega t}$
and $\Dabk=\epsilon_\alpha - \epsilon_\beta + k \Omega$.
The quantities $S_{\alpha\beta,k}$ and $\Dabk$ can be thought of as coupling
amplitudes and transition energies; the corresponding (partial) transition rates
are $a_{\alpha\beta,k} = |S_{\alpha \beta,k}|^2 s(\Dabk)$, where $s(\epsilon)=
\theta(\epsilon) J(\epsilon) n_B(\epsilon) + \theta(-\epsilon) J(-\epsilon)
\left[ n_B(-\epsilon)+1\right]$ and $\theta(\epsilon)$ is the Heaviside step
function.
The rates appearing in \eqref{eq:gqme_fsa} are obtained by summing the partial
rates over all $k$ contributions; namely, $ A_{\alpha\beta}^{\nu} = \sum_k
e^{-i \nu \Delta_{\alpha\beta,k}}  a_{\alpha\beta,k}$ and $A_{\alpha\beta}
\equiv A_{\alpha\beta}^{\nu = 0}$.
Taking the limit $\nu \to 0$ in (\ref{eq:gqme_fsa}), one recovers a
master equation found in literature, whose dynamic steady state (DSS, also known
as quasistationary state) is characterized by $z^{\st} \equiv
\rho_{22}^{\st}-\rho_{11}^{\st}=(A_{21}-A_{12})/(A_{12}+A_{21})$
\cite{Grifoni1998}.

Even before solving the GQME, we notice that the knowledge of
$\dot{\vec{\rho}}^\nu$ can be used to evaluate the mean heat power
$\mean{\dot Q}=d \mean{Q}/dt$ transferred to the reservoir at any time (see \ref{app:heat_power}):
\be
\mean{\dot Q} = - \sum_{\alpha,\beta,k} \Delta_{\alpha\beta,k} \;
a_{\alpha\beta,k} \; \rho_{\beta\beta} \ . \label{eq:Qdot} \ee 
An equivalent result was recently derived using the standard master equation approach and associating the relaxation processes to the dissipated heat \cite{LevyPRE2012, Szczygielski2013, langemeyer14}. 
Equation \eqref{eq:Qdot} can be used to classify open driven quantum
systems into two categories: those that exchange heat at steady-state, and those
that do not.
In general and in contrast to the undriven case, a driven system will keep
exchanging energy with the environment unless its dressed (Floquet) states are
decoupled (for instance, by symmetry) from the noise.

The GQME defined by \eqref{eq:gqme_fsa} has got an analytical solution. The
corresponding CF can be written as \be G(\nu,t)= c_-^\nu
e^{\xi_-(\nu) t} + c_+^\nu e^{\xi_+(\nu) t} \ ,
\label{eq:Glambda_sol} \ee where $\xi_\pm(\nu)$ are the eigenvalues of
$\mathcal{A}$ and $c_\pm^\nu$ the projection of the initial density matrix
$\rho(0)$ onto the corresponding eigenvectors $v_\pm^\nu$, normalized so
that $\Tr v_\pm^\nu = 1$.
Many properties of the heat distribution can be obtained from
$\eqref{eq:Glambda_sol}$; see \ref{app:results_long_time}.
We first notice that the first few moments of the distribution can be
straightforwardly calculated; in a rigorously defined long-time limit, the
cumulants $\mean{\mean{Q}}_n$ of the PDF \cite{Belzig2003} are all linear in
time and solely determined by the dominant eigenvalue $\xi_+(0)$:
\be
\mean{\mean{Q}}_n=(-i)^n \xi_+^{(n)}(0) t \ , \label{eq:moments}
\ee 
where $\xi_+^{(n)}$ is the $n$-th derivative of $\xi_+$ with respect to
$\nu$.
In the following we will consider the first three cumulants, which coincide with
the central moments, namely, mean value, variance and skewness of the PDF.

Furthermore, the explicit knowledge of the CF allows us to retrieve the
full PDF at any given time. This requires inverting the Fourier transform, a task which in general must be
performed numerically.
However, some analytical insight can be gained in the long-time limit.
We first notice that $\Real \left[ \xi_-(\nu) \right]$ is always negative, while $\Real \left[
\xi_+(\nu) \right]$ is negative everywhere except at $\nu_n = n\tau$, $n \in
\mathcal Z$, where it vanishes.
From \eqref{eq:Glambda_sol}, we thus see that $\lim_{t \to \infty} G(\nu,t) = 0$ for $\nu \neq \nu_n$.
This property suggests performing a series expansion around each $\nu_n$. After the expansion, the Fourier
transform can be inverted analytically.
In the following, we expand $\xi_+(\nu)$ up to the second order. This expansion
results in a Gaussian approximation for the PDF; accordingly, it correctly
reproduces the first two moments of the original distribution.
The PDF reads \be
\begin{split}
P(Q,t)=  w(Q,t) \Omega \sum_k \ \Big[  & p_\downarrow \delta(Q+\epsilon_1-\epsilon_2 + k\Omega)  \\
+\ & p_\uparrow \delta(Q+\epsilon_2-\epsilon_1 + k\Omega) \\
+\ & (1-p_\uparrow-p_\downarrow) \delta(Q + k\Omega) \Big] \ ,
\end{split}
\label{eq:PofE}
\ee
where $p_\uparrow = A_{21}/(A_{21}+A_{12}) \rho_{11}(0)$, $p_\downarrow = A_{12}/(A_{21}+A_{12}) \rho_{22}(0)$
and 
\be
w(Q,t) = \frac{1}{\sqrt{2\pi}} \frac{1}{\sqrt{2 b t}\Omega} \exp\left[-\frac{(Q - a t \Omega)^2}{4 b t \Omega^2}\right] \label{eq:w} \ ,
\ee
with real coefficients $a= -i \xi_+^{(1)}(0)\geq 0$ and $b= - 1/2 \xi_+^{(2)}(0) \geq 0$.

Equations \eqref{eq:PofE} and \eqref{eq:w} allow for an insightful interpretation of
the energy-exchange process.
The delta functions account for the fact that the exchanges take place only in
multiples of $\Delta_{\alpha\beta,k}$.
This discretization results from energy being available in photons of energy $\Omega$, the
drive frequency, and $\epsilon_2-\epsilon_1$, the dressed energy gap of the
driven system.
Furthermore, they tell us that the system is either found in the same Floquet state
or has undergone a transition upwards (downwards), with probability
$1-\pup-\pdn$ and $\pup$ ($\pdn$), respectively.
On the other hand, the total probability that a certain amount of energy has
been exchanged is dictated by the weight function $w(Q,t)$.
An appealing feature of \eqref{eq:PofE} is that it clearly shows how the
heat-exchange distribution ``builds up'' on individual exchanges of well-defined
energy.
One should keep in mind, however, that it was derived under the assumption that
$\CF$ is localized at the nodes $\nu= n \tau$.
This assumption holds true only in the limit where many energy quanta have been
exchanged.
For very long times, $t \gg 1/b$, one can neglect the discretization due to the
Dirac combs in \eqref{eq:PofE} and find that $P(Q,t) \approx w(Q,t)$.

\section{Examples}

\begin{figure}
\centering
\includegraphics[width=0.85\textwidth]{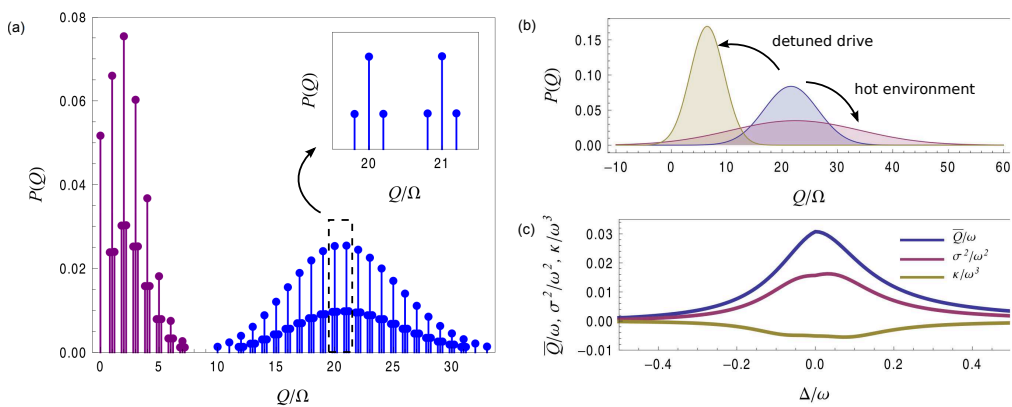}
\caption{
Dissipated energy for transverse coupling.
(a) PDF at $t/\tau=80$ (purple) and $t/\tau=700$ (blue).
Inset: triplets at frequencies $k \Omega$, and $k\Omega \pm \gap$.
(b) Effect of temperature and detuning on $w(Q,t)$.
Temperature is changed from $k_B T=0.1\omega$ to $k_B T=3\omega$. The detuning is changed from $\Delta=0.02\omega$ to $\Delta=0.3\omega$. 
(c) Mean value $\bar{Q}$ (blue), variance $\sigma^2$ (purple) and skewness $\kappa$ (yellow) of the PDF as a function of the detuning $\Delta$.
For all panels, the initial state is the DSS, $t/\tau=700$, $g=0.1\omega$, $\Delta=0.02\omega$, $\varphi = 0$, $k_B T = 0.1 \omega$ and $\eta = 0.01$.
}
\label{fig:sigma_x}
\end{figure}

The results discussed above are general and can be applied
to a two-level system with any drive and coupling operator to the environment.
As an illustrative, analytically tractable example, we discuss a two-level
system interacting with a monochromatic drive, described by the Rabi Hamiltonian
$H(t) = \omega/2~(1+\sigma_z) + g [e^{i (\Omega t-\varphi)} \sigma_+ + e^{-i
(\Omega t-\varphi)} \sigma_-] $ where $\sigma_i$ are the usual Pauli operators,
$\omega$ is the bare energy gap, $g$ the amplitude and $\varphi$ the phase of
the driving field.
The Floquet states can be written in analytic form (see \ref{app:Rabi_model}) and the
quasienergy gap $\epsilon_2-\epsilon_1$ equals the Rabi frequency $\Omega_R =
\sqrt{ \Delta ^2 +4 g^2}$, where $\Delta = \Omega-\omega$ is the drive detuning.
We consider two types of coupling operators, $S=\sigma_x$ (transverse coupling)
and $S=\sigma_z$ (longitudinal coupling).
Both can be realized, e.~g., in solid-state devices
\cite{Wilson2007,Wilson2010}.

When $S=\sigma_x$, \eqref{eq:Qdot} tells that the DSS heat current is, in
general, nonvanishing. In physical terms, the system is continuously ``pumped''
by the drive and emits photons to the environment, resulting in a net heat flow
to the environment.
In this case, the dependence of the PDF on the initial state is negligible at
long times. Therefore, we directly take the DSS as the initial state, for which
$p_\downarrow=p_\uparrow$.
The full PDF is shown in the main panel of figure \ref{fig:sigma_x} at times
$t_1/\tau = 80$ (purple) and $t_2/\tau = 700$ (blue).
It was obtained from the analytical solution \eqref{eq:Glambda_sol} by
numerically inverting the Fourier transform.
The structure of the PDF is the same as in \eqref{eq:PofE}: a series of Dirac
combs modulated by an envelope which moves in time.
As best seen in the Inset, the spectrum of possible energies is composed by
symmetric triplets centered at integer multiples of $\Omega$ and spaced by an
amount $\Omega_R$.
There is a strong analogy between this result and the Mollow triplet
\cite{Mollow1969} observed in quantum-optics experiments
\cite{Cohen-Tannoudji1998,Haroche2006}.
As for the envelope function of the PDF, it is manifestly non-Gaussian at early
times.
In the long-time limit, the PDF tends to a Gaussian as the relative importance of higher-order cumulants decreases [see equation \eqref{eq:moments}].

In figure \ref{fig:sigma_x} (b,c), we show how the PDF is affected by changes in the drive frequency and the temperature of the environment. 
We take a long time ($t/\tau=700$) and use the analytical results \eqref{eq:w} and \eqref{eq:moments}.
In panel (b), we plot the Gaussian envelope function $w(Q,t)$ for three representative
cases.
In passing from low to high temperature, the mean value of the distribution is barely affected, while the
variance strongly increases.
The fact that temperature has little influence on the average dissipated heat is
related to the fact that -- differently from an undriven system -- the
transition rates $A_{\alpha\beta}$ do not satisfy the detailed balance.
On the contrary, the variance of the distribution depends on temperature, due to the
fact that a higher-temperature environment leads to stronger noise effects.
In passing from a resonant to a red-detuned drive, the PDF shows reduced average and variance because of the
reduction of the energy injected in the system that can then be dissipated.
This trend is confirmed by the behaviour of the central moments of the PDF as a function of
$\Delta$, shown in panel (c). The maxima of the moments are reached at resonance, i.e., when $\Delta=0$.

\begin{figure}
\centering
\includegraphics[width=0.7\linewidth]{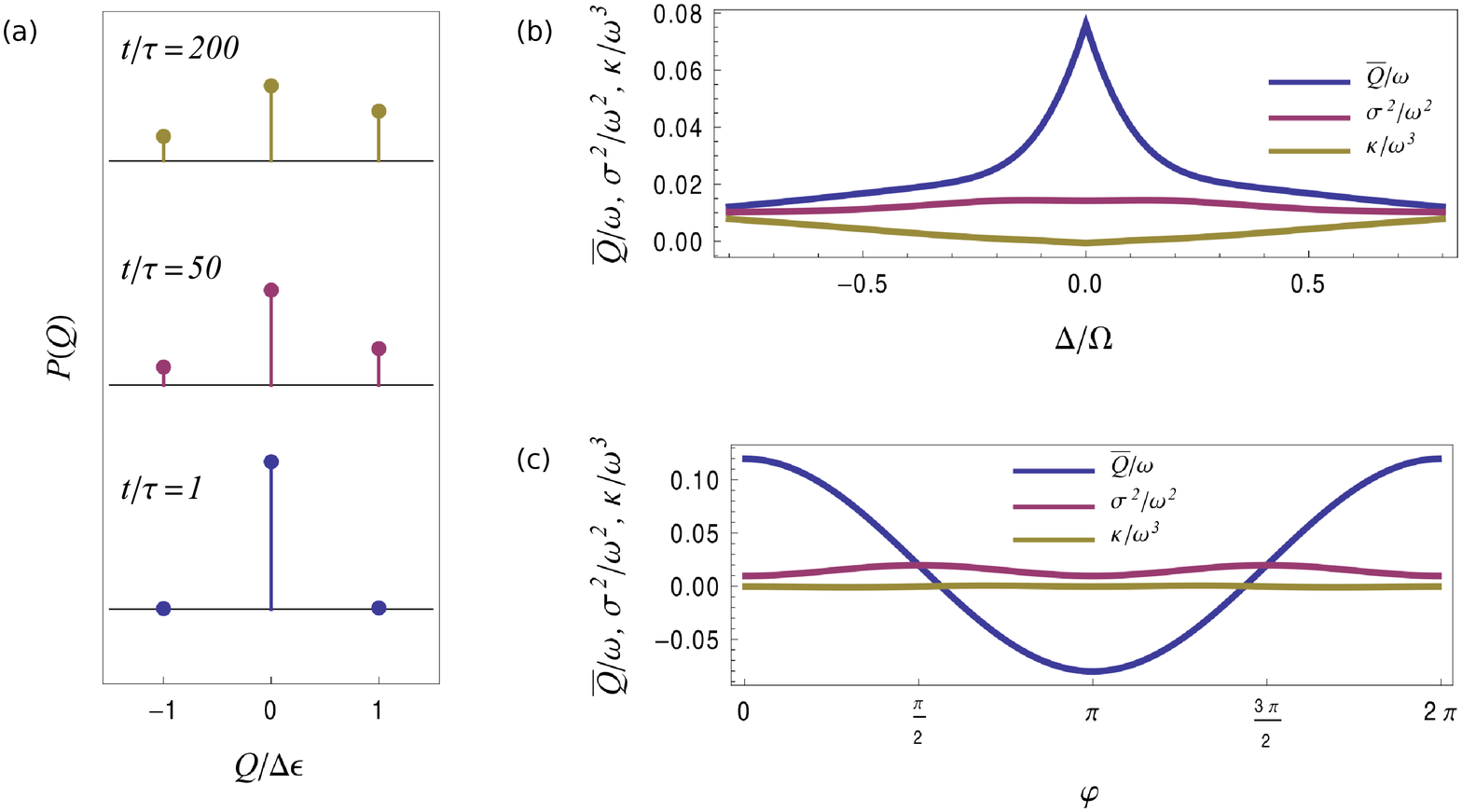}
\caption{
Dissipated energy for longitudinal coupling.
(a) PDF at different times. (b, c)
First three central moments of the dissipated heat as a function of the detuning $\Delta$ (b) and the drive phase $\varphi$ (c).
The initial state is $\ket{0}$ in (a, b) and $\frac{1}{\sqrt{2}}\left(\ket{0}+\ket{1}\right)$ in (c).
A high temperature $k_BT/\omega=0.5$ is considered in (c). All other parameters are as in figure \ref{fig:sigma_x}.
}
\label{fig:sigma_z}
\end{figure}

We next consider the noise operator $S= \sigma_z$.
In this case, \eqref{eq:Qdot} tells that the mean heat power vanishes at DSS.
Therefore, the heat depends critically on the initial state since it is dissipated before the reaching of the DSS.
The absence of a net heat flow at DSS is due the symmetry of the problem, which forbids transitions with energy exchange $n\Omega$ and thereby
causes the drive and the environment to be effectively decoupled \cite{langemeyer14}.
The resulting GQME is formally equivalent to that for an undriven system; its derivation can be found in \cite{app:GQME_time_indep}. The corresponding PDF reads
 \be
P(Q,t) = p_\downarrow(t) \delta(Q- \gap) + p_\uparrow(t) \delta(Q+\gap) + p_0(t) \delta(Q) \ ,
\ee
where
$ p_\downarrow(t) = \Gamma_-/\Gamma \left(1-e^{- \Gamma t}\right) \rho_{22}(0)  $, 
$p_\uparrow(t) =   \Gamma_+/\Gamma \left(1-e^{- \Gamma t}\right) \rho_{11}(0) $ and
$p_0(t) = 1-p_\downarrow(t)-p_\uparrow(t)$,  $\Gamma_\pm =  2\pi  S_{12,0}^2 s(\pm \gap)$ and $\Gamma =  \Gamma_+ + \Gamma_-$.
Notice that the structure of this PDF is the same as the more general  (\ref{eq:PofE}),
with weight function $w(E,t) = 1$ for $-\frac{\Omega}{2} < E < \frac{\Omega}{2}$ and $0$ otherwise.
Figure \ref{fig:sigma_z} (a) shows a typical time evolution of the PDF as the
system reaches its DSS.
Figure \ref{fig:sigma_z} (b) shows the dependence of the first three central
moments on the detuning $\Delta$, starting from the ground state of the undriven
system.
If, by contrast, the initial state is in a coherent superposition of the undriven system eigenstates
$\ket{0}$ and $\ket{1}$, then the drive phase $\varphi$ also affects the heat
distribution, depending on the projection of the initial state onto the
$\hat{xy}$ plane.
An example is shown in figure \ref{fig:sigma_z} (c) for an equal superposition of $\ket{0}$ and $\ket{1}$.
Notably, there exists a range of values where the mean exchanged heat becomes
negative, meaning that the environment is more likely to emit energy and the
system to absorb it.

\section{Conclusions} We have developed a general formalism to study the
distribution of heat exchanges between a periodically driven quantum system and
a heat bath.
Our approach, based on a combination of Floquet theory and generalized quantum
master equation, fully takes into account the effects of the drive.
Its power is epitomized by the possibility of obtaining the full heat-exchange
distribution, as in Eqs.~\eqref{eq:PofE} and \eqref{eq:w}.

Some of the assumptions made in this work could be further relaxed.
The full secular approximation may be replaced by a partial secular
approximation, which gives more accurate results in the vicinity of quasienergy
crossings \cite{Grifoni1998,Gasparinetti2013}.
Several techniques developed in the context of the full-counting statistics may
be readily adapted to the present setting, including weak-coupling Markovian
approaches \cite{Bagrets2003,Flindt2005}, non-Markovian perturbative expansion
\cite{Braggio2006}, recursive schemes \cite{Flindt2008,Flindt2010} and
waiting-time analysis \cite{Brandes2008,Albert2011,Thomas2013}.
These techniques, originally developed for undriven systems, can find
application here as the Floquet picture turns a time-dependent problem into a
time-independent one.

Our predictions could be tested in a variety of physical systems, for
example, superconducting quantum bits embedded in a resistive environment
\cite{Solinas2013,Pekola2013} and/or integrated in a
circuit-quantum-electrodynamics architecture \cite{Wallraff2004,Murch2012}.
Their experimental confirmation would improve our current understanding of the
thermodynamics of driven quantum systems and may as well open new avenues for
the realization of thermal machines based on quantum mechanics.

\textbf{Acknowledgements} We would like to thank J.~Pekola and R.~Fazio for a
careful reading of the manuscript.
This work has received funding from the European Union FP7/2007-2013 under REA
grant agreement no 630925 -- COHEAT, from MIUR-FIRB2012 -- Project HybridNanoDev
(Grant No.~RBFR1236VV) and from MIUR-FIRB2013 -- Project Coca (Grant
No.~RBFR1379UX). S.G. acknowledges financial support from the Finnish National
Graduate School in Nanoscience (NGS-NANO) and the Network in Condensed Matter
and Materials Physics (CMMP).

\section*{References}

\bibliographystyle{iopart-num}
\bibliography{Biblio-Quantum_Calor}

\appendix

\section{Derivation of the Generalized Quantum Master Equation}
\label{app:MEQ_derivation}

Starting from the definition of $\CF$ discussed in the main text, we can write a generalized quantum master equation (GQME) for the system generalized density matrix $\rho^\nu =\Tr_R [\rho_T^\nu]$, in analogy with the usual quantum master equation \cite{Breuer2007}.
The standard master equation is recovered in the limit $\nu=0$.

We start from the equation of motion for the total generalized density operator $\rho_T^\nu(t)$ in (\ref{eq:rho_T_def})
\be \dot \rho_T^\nu(t) =
-i[H(t)+H_R,\rhoT^\nu]-i (\Vlambda \rhoT^\nu - \rhoT^\nu \Vmlambda) \ ,
\ee where $V_\nu=e^{i\nu H_R}Ve^{-i\nu H_R}= S \otimes B_\nu = S \otimes e^{i\nu H_R}Be^{-i\nu H_R}$. We work in the interaction picture defined by $H(t)+H_R$, so that
a generic operator $O$ is transformed as $O^{(I)}
=e^{i H_R t} U_0^\dagger(t)O U_0(t) e^{-i H_R t}$,
where $U_0(t)$ is the evolution operator generated by $H(t)$.
After a simplification, we get \be \dot{\rho}_T^{\nu, (I)}(t) = -i
(\Vlambda^{(I)} \rhoT^{\nu, (I)} - \rhoT^{\nu, (I)} \Vmlambda^{(I)}) \ .
\label{eq:rholambdadot} \ee
We then perform a series expansion up to the second order in the
system-environment coupling and invoke the Born approximation, which implies
that $\rhoI_T(t)=\rhoI(t) \otimes \rho_R(0)$ \cite{Breuer2007}.
We introduce a generalized correlation function \be g_{\nu}(t'-t'') =
\Tr_R \left[ B_{-\frac{\nu}{2}}^{(I)}(t') \rhoI_R
B_{\frac{\nu}{2}}^{(I)}(t'') \right] \ . \ee
Its explicit evaluation gives
\be
g_\nu(\tau) = \sum_{n,n'} e^{i (e_n-e_{n'})(\tau+\nu)} B_{nn'} \label{eq:gnu_eval}
B_{n'n} p_{n} \ ,
\ee where $e_n$ are the eigenenergies of the bath and
$p_n=e^{-\beta e_n}/Z_R$ the corresponding occupation probabilities. From \eqref{eq:gnu_eval} we see that
\be{ g_\nu(\tau)=g_0(\tau+\nu) \equiv g(\tau+\nu)} \ , \ee
implying that the generalized correlation function $g_\nu$ is a time-shifted
ordinary correlation function.
For a bosonic bath, the correlation function takes the form \be
\begin{split}
g(\tau) = &  \int_0^\infty d\omega J(\omega) \left[ e^{i\omega\tau} n_B(\omega) + e^{-i\omega\tau} \left( n_B(\omega)+1\right)  \right] \\
= & \int_0^\infty d\omega J(\omega) \left[ \cos(\omega \tau) \coth(\beta\omega/2) - i \sin (\omega \tau) \right]
\ , \end{split} 
\ee
where $J(\omega)$ is the spectral density associated to the coupling operator $B$ and $n_B(\omega)$ the Bose distribution function.
Performing the Born-Markov approximation and extending the integration times to infinity, we obtain 
\begin{equation}
\begin{split}
\frac{\partial}{\partial t} \rho^{\nu, (I)} = - & \int_0^\infty d\tau \\
& g(\tau) S^{(I)}(t) S^{(I)}(t-\tau) \rho^{\nu, (I)} \\ 
- & g(-\nu+\tau) S^{(I)}(t-\tau) \rho^{\nu, (I)} S^{(I)}(t) \\
- & g(-\nu-\tau) S^{(I)}(t)  \rho^{\nu, (I)} S^{(I)}(t-\tau) \\ 
+ & g(-\tau) \rho^{\nu, (I)} S^{(I)}(t-\tau)  S^{(I)}(t) \ ,
\end{split}
 \label{eq:gqme_int}
\end{equation}
Sometimes it is convenient to go back to the Schr\"odinger picture. In this case we have:
\begin{equation}
\begin{split}
\frac{\partial}{\partial t} \rho^\nu = & - i \left[ H, \rho^\nu \right]  
	-  \int_0^\infty d\tau \\
& g(\tau) S(0) S(-\tau) \rho^\nu \\ 
- & g(-\nu+\tau) S(-\tau) \rho^\nu S(0) \\
- & g(-\nu-\tau) S(0)  \rho^\nu S(-\tau) \\ 
+ & g(-\tau) \rho^\nu S(-\tau)  S(0) \ .
\end{split} \label{eq:gqme_schro}
\end{equation}
where $S(-\tau)$ stands for $U_0^\dagger (t-\tau,t)S U_0(t-\tau,t)$.

\section{GQME for a periodic drive}
\label{app:GQME_periodic}

We consider the case of a periodic drive, so that $H(t+\tau)=H(t)$ with
$\tau=2\pi/\Omega$. The dynamics of the driven system is captured by considering
the Floquet states $\ket{\Psi_\alpha(t)}$ and their corresponding quasienergies
$\epsilon_\alpha$. We also introduce the Floquet modes $\ket{\phi_\alpha(t)}$,
satisfying $\ket{\phi_\alpha(t+\tau)}=\ket{\phi_\alpha(t)}$ and related to the
Floquet states by $\ket{\Psi_\alpha(t)}=e^{-i \epsilon_\alpha
t}\ket{\phi_\alpha(t)}$.

The evolution operator of the system in the Floquet basis reads $U_0=\sum_\alpha \ketbra{\Psi_\alpha(t)}{\Psi_\alpha(0)}$.
We work in the interaction picture defined by $U_0(t)$.
Then
\begin{equation}
\begin{split}
S^{(I)}(t) &= \sum_{\alpha,\beta}\ketbra{\Psi_\alpha(0)}{\Psi_\alpha(t)} S \ketbra{\Psi_\beta(t)}{\Psi_\beta(0)}  \\
&= \sum_{\alpha,\beta,k} S_{\alpha\beta,k}e^{ i \Delta_{\alpha\beta,k} t} \ketbra{\Psi_\alpha(0)}{\Psi_\beta(0)} \ ,
\end{split}
\end{equation}
with
\begin{align}
S_{\alpha\beta,k} &= \frac{1}{\tau} \int_0^\tau dt \xbraket{\phi_\alpha(t)}{S}{\phi_\beta(t)} e^{- i k \Omega t} \label{eq:S_al,be,k} \ , \\ 
\Delta_{\alpha\beta,k} &= \epsilon_\alpha - \epsilon_\beta + k \Omega \ .
\end{align}
Notice the property $S_{\alpha\beta,k} = S_{\beta \alpha, -k}^*$.

By substituting \eqref{eq:S_al,be,k} into \eqref{eq:gqme_int}, and denoting with $\rho_{\alpha\beta}^\nu= \xbraket{\Psi_\alpha}{\rho^\nu}{\Psi_\beta}$, we get
\be
\begin{split}
\dot\rho_{\alpha\beta}^{\nu, (I)} = - & \sum_{\gamma,\delta,k,k'} \int_0^\infty d\tau \\
& g(\tau) S_{\alpha\gamma,k}e^{ i \Delta_{\alpha\gamma,k} t} S_{\gamma\delta,k'} e^{ i \Delta_{\gamma\delta,k'} (t-\tau)} \rho_{\delta\beta}^{\nu, (I)} \\ 
- & g(-\nu+\tau) S_{\alpha\gamma,k}e^{ i \Delta_{\alpha\gamma,k} (t-\tau)} \rho_{\gamma\delta}^{\nu, (I)} S_{\delta\beta,k'}e^{ i \Delta_{\delta\beta,k'} t} \\
- & g(-\nu-\tau) S_{\alpha\gamma,k}e^{ i \Delta_{\alpha\gamma,k} t}  \rho_{\gamma\delta}^{\nu, (I)} S_{\delta\beta,k'}e^{ i \Delta_{\delta\beta,k'} (t-\tau)} \\ 
+ & g(-\tau) \rho_{\alpha\gamma}^{\nu, (I)} S_{\gamma\delta,k}e^{ i \Delta_{\gamma\delta,k} (t-\tau)}  S_{\delta\beta,k'}e^{ i \Delta_{\delta\beta,k'} t} \ .
\end{split}  \label{eq:gqme_001}
\ee
Equation \eqref{eq:gqme_001} can be rewritten as:
\be
\begin{split}
\dot\rho_{\alpha\beta}^{\nu, (I)} = -& \sum_{\gamma,\delta,k,k'} 
 I_+(0,\Delta_{\gamma\delta,k'}) S_{\alpha\gamma,k} S_{\gamma\delta,k'}   e^{ i (\Delta_{\alpha\gamma,k}  + \Delta_{\gamma\delta,k'}) t} \rho_{\delta\beta}^{\nu, (I)}  \\ 
- & I_+(-\nu,\Delta_{\alpha\gamma,k}) S_{\alpha\gamma,k}  S_{\delta\beta,k'}  e^{ i (\Delta_{\alpha\gamma,k}  +  \Delta_{\delta\beta,k'}) t} \rho_{\gamma\delta}^{\nu, (I)} \\
- & I_-(-\nu,\Delta_{\delta\beta,k'}) S_{\alpha\gamma,k}  S_{\delta\beta,k'} e^{ i (\Delta_{\alpha\gamma,k} +  \Delta_{\delta\beta,k'}) t} \rho_{\gamma\delta}^{\nu, (I)} \\ 
+ & I_-(0,\Delta_{\gamma\delta,k})  S_{\gamma\delta,k}  S_{\delta\beta,k'}  e^{ i(\Delta_{\gamma\delta,k}  +  \Delta_{\delta\beta,k'}) t} \rho_{\alpha\gamma}^{\nu, (I)} \ ,
\end{split} \label{eq:general_GQME}
\ee
where we have introduced the integrals of the correlation function
\be
I_\pm(\nu,E)=\int_0^\infty d\tau g(\nu \pm \tau) e^{-i E \tau} \ . \label{eq:integrals_def}
\ee
The integrals \eqref{eq:integrals_def} can be explicitly evaluated:
\be
\begin{split}
I_\pm(\nu,E)= &\int_0^\infty d\tau e^{-i E \tau}
\int_0^\infty d\omega \\
J(\omega) &\left[ e^{i\omega\nu}e^{\pm i\omega\tau} n_B(\omega) 
+ e^{-i\omega\nu}e^{\mp i\omega\tau} \left( n_B(\omega)+1\right)  \right] \ . 
\end{split}
\ee
Using the identity $\int_0^\infty e^{i\omega t} dt = \pi \delta(\omega)+
\mathcal{P} (\frac{i}{ \omega})$ and neglecting principal-value integrals, we obtain 
\be
 {
I_\pm(\nu,E)= \pi e^{\pm i E \nu} s(\pm E) \ ,
} 
\label{eq:g_int}
\ee
where
\be
s(E)=\theta(E) J(E) n_B(E) + \theta(-E) J(-E) \left[ n_B(-E)+1\right]
\label{eq:spectral_density}
\ee
and $\theta(E)$ is the Heaviside step function.
 
Our results so far apply to \emph{any} periodically driven system.
In the following, we will take the fast-drive limit.

\subsection*{GQME in the full secular approximation}

We assume $\Omega$ to be faster than the decoherence rates that appear in the
GQME. This allows us to neglect fast-oscillating terms of the form $e^{i k
\Omega t}$, $k\neq 0$. A further class of terms, oscillating as $e^{ i
(\epsilon_\alpha - \epsilon_\beta) t}$, $\alpha\neq\beta$, can be safely
neglected in most \cite{Grifoni1998, Russomanno2011}, but not all
\cite{Gasparinetti2013}, instances.
Altogether, these two approximations amount to a full secular approximation,
which results in a decoupling of the equations for the diagonal and off-diagonal
elements in \eqref{eq:general_GQME}.
We note in passing that in this approximation, the principal-value integrals that we neglected in deriving \eqref{eq:g_int} would not contribute to $G(\nu,t)$.

In order to determine $\CF$, it is enough to solve \eqref{eq:general_GQME} for the diagonal elements $\rho_{ii}^\nu$.
We define
\be
\begin{split}
a_{\alpha\beta,k} &= 2 \pi s(\Delta_{\alpha\beta,k}) |S_{\alpha \beta,k}|^2  \ , \\
A_{\alpha\beta} &= \sum_k a_{\alpha\beta,k}  \ , \\
A_{\alpha\beta}^{\nu} &= \sum_k e^{-i \nu \Delta_{\alpha\beta,k}}  a_{\alpha\beta,k} \ .
\label{eq:a_defs}
\end{split}
\ee

At this point we focus on a two-level system, which simplifies the analytical treatment.
The GQME \eqref{eq:general_GQME} can be written as:
\be
\begin{split}
\dot\rho_{11}^\nu&= [A_{11}^\nu-A_{11}-A_{21}]\rho_{11}^\nu + A_{12}^{\nu}\rho_{22}^\nu  \ , \\
\dot\rho_{22}^\nu &= A_{21}^{\nu}\rho_{11}^\nu + [A_{22}^\nu-A_{22}-A_{12}]\rho_{22}^\nu \ .
\end{split}  \label{eq:fsa_gqme}
\ee
We can also write \eqref{eq:fsa_gqme} in matrix form by introducing a vector $\vec{\rho^\nu} = \{\rho_{11}^\nu, \rho_{22}^\nu\}$ and a matrix $\mathcal{A}$ so that
\be
\dot{\vec{\rho^\nu}} = \mathcal{A} \cdot \vec{\rho^\nu} \ . \label{eq:gqme_vec}
\ee
In the limit $\nu \to 0$, one recovers the standard master equation
\be
\begin{split}
\dot\rho_{11} &= -A_{21}\rho_{11} + A_{12}\rho_{22} \ ,  \\
\dot\rho_{22} &= A_{21}\rho_{11} - A_{12}\rho_{22}  \ ,
\end{split} \label{eq:fsa_qme}
\ee
whose stationary solutions or dynamical steady state (DSS) are the well known \cite{Grifoni1998}
\be
\begin{split}
\rho_{11}^{\rm st} &= \frac{A_{12}}{A_{12}+A_{21}} \ ,  \\
\rho_{22}^{\rm st} &= \frac{A_{21}}{A_{12}+A_{21}} \ . 
\end{split} \label{eq:fsa_sol}
\ee
Notice that for a two-level system any noise operator $S$ is traceless, so that $ S_{22}=- S_{11}$.
As a result, the coefficients appearing in \eqref{eq:a_defs} and \eqref{eq:fsa_gqme} satisfy the relation $a_{22,k} = a_{11,k} $, which in turn implies $A_{22} = A_{11} $ and $A_{22}^{\nu} = A_{11}^{\nu} $.

\section{Average heat power}
\label{app:heat_power}

Let us calculate the time variation of the average heat, which we refer to as the mean heat power. It is given by 
\be
\mean{\dot Q} = \frac{d}{dt} \mean{Q}= -i \left. \frac{d}{d t} \frac{d}{d\nu} \CF \right|_{\nu = 0} = -i \left. \frac{d}{d\nu} \dot G(\nu,t) \right|_{\nu = 0}  \ .
\ee
From the definition of the CF we have that $\CF = \Tr_S [ \rho^\nu] =  \rho_{11}^\nu+\rho_{22}^\nu$.
The evolution of the generating function is given by $\dot G(\nu,t) = \dot\rho_{11}^\nu+\dot\rho_{22}^\nu$.
Using \eqref{eq:fsa_gqme} we obtain
\be  {
\mean{\dot Q} = - \sum_{\alpha,\beta,k} \Delta_{\alpha\beta,k} \; a_{\alpha\beta,k} \; \rho_{\beta\beta} } \ .
\label{eq:dQ_app}
\ee
This equation directly relates the dissipated heat power to the dynamics of the system.
Given a solution $\rho$ of the standard master equation, which is in general easier to obtain than the one for the GQME, we can immediately predict the dissipated power using \eqref{eq:dQ_app}.
The most interesting application is in determining if and how much heat is dissipated when the system reaches the DDS \eqref{eq:fsa_sol}.
As discussed in the main text and in the following examples, this allows us to classify the driven systems in dissipative or non-dissipative at steady-state.

\section{General results in the long-time limit}
\label{app:results_long_time}

We define
\be
\begin{split}
\Sigma &=A_{12}+A_{21} \ , \\
\Upsilon_{\alpha\beta}(\nu) &= A_{\alpha\beta}^\nu-A_{\alpha\beta} \ , \\
\Upsilon_{\alpha\beta}^{(2)}(\nu) &= A_{\alpha\beta}^\nu A_{\beta\alpha}^\nu-A_{\alpha\beta}A_{\beta\alpha}  \ . \label{eq:upsilonab}
\end{split}
\ee
The eigenvalues of $\mathcal{A}$ are
\be
\xi_\pm(\nu) = \frac12 \left(r \pm h \right) 
\ee 
with
\begin{align}
h &= \sqrt{\Sigma^2+4 \Upsilon_{12}^{(2)}(\nu)} \ ,  \\
r &= - \Sigma + \Upsilon_{11}(\nu) +\Upsilon_{22}(\nu)   \ .
\end{align}
The corresponding eigenvectors are
\begin{align}
v_\pm^\nu &= \frac{1}{A_{12}-A_{21}+2A_{21}^\nu\pm h} \vectwo{A_{12}-A_{21}\pm h}{ 2 A_{21}^\nu}. 
\end{align}
Notice that in this representation the trace of a density operator is obtained by summing the two vector components. 
In the above equation, we have chosen the normalization constant for $v_\pm^\nu$ so that $\Tr[v_\pm^\nu]=1$.

The initial density matrix can be characterized by the quantity $z= \rho_{22}(0) - \rho_{11}(0)$. We can write it as
\be
\rho(0)=\begin{pmatrix}{\rho_{11}(0)}\\{\rho_{22}(0)}\end{pmatrix}= \begin{pmatrix}{\frac{1-z}2}\\{\frac{1+z}{2}}\end{pmatrix} = c_+^\nu v_+^\nu + c^\nu_- v_-^\nu 
\ee
with
\begin{align}
c_-^\nu &= \frac{1}{2 h } \left[ (h - A_{12}^\nu - A_{21}^\nu) + (\Upsilon_{21}(\nu)-\Upsilon_{12}(\nu)) z \right]  \ , \\
c_+^\nu &= \frac{1}{2 h} \left[ (h + A_{12}^\nu + A_{21}^\nu) + (\Upsilon_{12}(\nu) - \Upsilon_{21}(\nu)) z \right] \ .
\label{eq:c_lambda}
\end{align}
Putting things together, we finally write the CF as
\be
G(\nu,t)= c_-^\nu e^{\xi_-(\nu) t} + c_+^\nu e^{\xi_+(\nu) t}\ . 
 \label{eq:Glambda_eigen}
\ee
Equation~\eqref{eq:Glambda_eigen} is the exact solution of the GQME \eqref{eq:fsa_gqme}.
It can be used to calculate all the moments and cumulants of the PDF. 
In general, in order to obtain the full PDF, we must numerically invert the Fourier transform in \eqref{eq:G_def}.
In the following, however, we show that some analytical insight can be gained in the long-time limit.

\subsection*{Analytical expressions for the first moments}

The moments $\average{Q^n(t)} = (-i)^n \left.\frac{d^n G(\nu,t)}{d\nu^n}\right|_{\nu=0}$ can be calculated directly from \eqref{eq:Glambda_eigen}. 
From a physical point of view, the central moments $\average{(\delta Q)^n}$, with $\delta Q = Q- \average{Q}$, are the most meaningful quantities.
For $n=1, 2, 3$, these coincide with the cumulants $\mean{\mean{Q}}_n$ \cite{Belzig2003}.
In the long-time limit, the cumulants can be calculated directly from the dominant eigenvalues $\xi_+(\nu)$, as 
\be
 \mean{\mean{Q}}_n = (-i)^n t \frac{\partial^n}{\partial \nu^n} \xi_+(\nu) \Big|_{\nu=0} = (-i)^n \xi_+^{(n)}(0) t \ .
\ee
This is possible because at $\nu=0$, $\xi_+(0)=0$ and $\xi_-(0)=-\Sigma < 0$. As a result, all the terms involving derivatives of $\xi_-$ undergo exponential decay.

A PDF is often characterized by considering its first three central moments, namely, the
mean value $\bar{Q} = \average{Q}$, the variance
$\sigma^2=\average{(\delta Q)^2}$, and the skewness $\kappa=
\average{(\delta Q)^3}$.
In the long time limit, all these quantities increase linearly in time and are proportional to the derivatives of the dominant eigenvalue $\xi_+$ at $\nu=0$:
\begin{align}
\bar{Q}  &= - i \xi_+^{(1)}(0) t \ ,  \\ 
\sigma^2 &= - \xi_+^{(2)}(0) t \ , \\
\kappa &=i \xi_+^{(3)}(0) t  \ .
\end{align}

\subsection*{Eigenvalue analysis}

Using \eqref{eq:a_defs} and \eqref{eq:upsilonab}, $\Upsilon_{\alpha \alpha}(\nu)$ can be rewritten as (for $\nu \in \mathbb{R}$)
\be
\Upsilon_{\alpha \alpha}(\nu) = \sum_k (e^{-i\nu k \Omega} -1)a_{\alpha \alpha,k} \ ,
\ee
so that $\Real[\Upsilon_{\alpha \alpha}(\nu)] \leq 0$. Likewise,
\be
\Upsilon_{\alpha \beta}^{(2)}(\nu) = \sum_{k,k'} (e^{-i\nu (k+k')\Omega} -1)a_{\alpha \beta,k}a_{\beta \alpha,k'}  \ , \label{eq:Deltaxy2}
\ee
so that also $\Real[\Upsilon_{\alpha \beta}^{(2)}(\nu)] \leq 0$.

Altogether, this implies that $\Real \ r \leq -(A_{12}+A_{21})$ and $\Real \ h \leq
(A_{12}+A_{21})$. This implies that $\Real \ \xi_- < 0$, $\Real \ \xi_+ \leq 0$ and
\be
 \Real \ \xi_+ = 0~{\rm for}~ \nu = n \tau\ , \; n \in \mathcal{Z} \ .
\ee
Notice that $\xi_\pm(\nu)$ are periodic with period $\tau$, i.e.,
$\xi_\pm(\nu+n\tau)=\xi_\pm(\nu)$ and that for the subdominant
eigenvalues ${\rm max}[\Real \ \xi_-]=-\Sigma$ and the maximum is reached for $\nu = n
\tau$.
This means that the terms associated to the eigenvalue $\xi_-$ vanish
exponentially as soon as $t \gg 1/{\rm min}[\Real \ \xi_-]=1/\Sigma$ and the
behaviour of $\CF$ is determined only by $\xi_+$.
Furthermore, due to the fact that $\Real \ \xi_+ \leq 0$ for $\nu \neq n \tau$,
$\CF$ is different for zero only in a neighborhood of the resonances $\nu = n \tau$ and
vanishes exponentially away from them.

\subsection*{Solution in the long-time limit}
\label{app:long_time_limit}

In the long-time limit, $\CF \approx c_+^\nu e^{\xi_+(\nu) t}$ vanishes
everywhere except in a neighborhood of $\nu_n = n\tau$.
We recall that $\xi_+$ is periodic in $\tau$. In a neighborhood of $\nu_n$, we perform a second-order (Gaussian) approximation and write
\be
\xi_+(\nu) \approx i (\nu
\Omega - 2 n \pi) a - (\nu\Omega- 2 n \pi)^ 2 b \ ,
\label{eq:xipl_exp}
\ee
where $a = -i \xi_+^{(1)}(0)$ and $b=-1/2 \xi_+^{(2)}(0)$ with $a \leq 0$ and $b >0$.
Then we approximate the CF as
\be
G(\nu,t)= \sum_{n=-\infty}^\infty G_n e^{i (\nu  \Omega - 2 n
\pi) a t - (\nu\Omega- 2 n \pi)^ 2b t} \ ,
\label{eq:G_with_nodes}
\ee
where $G_n = c_+^{\nu= n\tau}$. Explicitly:
\be
\begin{split}
G_n &= 1+\frac{\Upsilon_{12}(n\tau)+\Upsilon_{21}(n\tau) + [\Upsilon_{12}(n\tau) - \Upsilon_{21}(n\tau)] z}{2(A_{12}+A_{21})}  \\
&= 1+ \frac{\rho_{11}(0) \Upsilon_{21} (n\tau)+\rho_{22}(0) \Upsilon_{12} (n\tau)}{A_{12}+A_{21}} \ . \label{eq:Gn}
\end{split}
\ee
Using \eqref{eq:a_defs} and \eqref{eq:upsilonab}, we can write
\be
\Upsilon_{\alpha \beta} (n\tau)= \left[ e^{-i n (\epsilon_\alpha-\epsilon_\beta) \tau}-1 \right] A_{\alpha \beta} \ . \label{eq:Upsilonn}
\ee
The Fourier transform in \eqref{eq:G_def} can be inverted analytically starting from \eqref{eq:G_with_nodes}.
The result is
\be
P(Q, t) = w(Q, t) \sum_{n=-\infty}^\infty G_n e^{- i n Q \tau} \ ,
\ee
where we have introduced the weight function
\be  {
w(Q, t) = \frac{1}{\sqrt{2\pi}} \frac{1}{\sqrt{2 b t}\Omega} \exp\left[-\frac{(Q - a t \Omega)^2}{4 b t \Omega^2}\right] } \ . 
\ee
Making use of \eqref{eq:Gn} and \eqref{eq:Upsilonn}, we obtain 
\be
P(Q, t) = w(Q, t) \sum_{n=-\infty}^\infty \Big[p_\downarrow e^{-i n (Q+\epsilon_1-\epsilon_2) \tau}  
+p_\uparrow e^{-i n (Q+\epsilon_2-\epsilon_1) \tau}
+ (1- p_\uparrow - p_\downarrow) e^{-i n Q \tau} \Big] \ , \label{eq:partial101}
\ee
where
\be
\begin{split}
 p_\uparrow &= \frac{A_{21} \rho_{11}(0)}{A_{12}+A_{21}} \ , \\
  p_\downarrow &= \frac{A_{12} \rho_{22}(0)}{A_{12}+A_{21}} \ . 
\end{split}
\ee
Now we recall the identity (Dirac comb)
\be
\sum_n e^{- i n Q \tau} = \Omega \sum_n \delta(Q+n\Omega) \ ,  \label{eq:DiracComb}
\ee
with $\tau= 2 \pi/\Omega$. Substituting \eqref{eq:DiracComb} into \eqref{eq:partial101}, we finally get
\be
\begin{split}
P(Q, t) =  w(Q, t) \Omega  \sum_{n=-\infty}^\infty &\Big[p_\downarrow \delta(Q+\epsilon_1-\epsilon_2 + n\Omega) 
+p_\uparrow \delta(Q+\epsilon_2-\epsilon_1 + n\Omega) \\
&+ (1- p_\uparrow - p_\downarrow) \delta(Q+ n\Omega)\Big] \ .
\end{split}
\label{eq:PofE_app}
\ee
The analytical result \eqref{eq:PofE_app} stems from the Gaussian expansion \eqref{eq:xipl_exp} and
is valid in the long-time limit.
This limit is defined by the condition $t \gg 1/b$, for which the CF is
localized at the nodes $\nu= n \tau$.
This is also the limit where many photons have been exchanged, as the standard
deviation of the Gaussian function $2 \Omega \sqrt{bt} $ is much larger than the
photon energy $\Omega$.
In this limit, the delta functions in \eqref{eq:PofE_app} tend to a
continuum and, as a result, we can write $P(Q,t) \approx w(Q,t)$.
The coefficient $a$ and $b$ are then directly linked to the average exchanged heat $\bar{Q} = \Omega a t$ and
variance $\sigma^2 = 4 \Omega^2 b t$, as expected from the analysis of the central moments.

\section{Two-level system with a monochromatic drive}
\label{app:Rabi_model}
Let us consider a two-level system driven by a transverse monochromatic drive.
This is also referred to as the Rabi model, described by the Hamiltonian 
\begin{equation}
 H(t) = \frac{ \omega}{2} (1+\sigma_z) + g \Big[e^{i (\Omega t-\varphi)} \sigma_+ + e^{-i (\Omega t-\varphi)} \sigma_-\Big] \ , 
\end{equation}
where $\sigma_i$ are the usual Pauli operators, defined with respect to the fixed basis $\{ \ket{0}, \ket{1} \}$. 
The Floquet modes read
\be
\begin{split}
 \ket{\phi_1} &= \cos \theta \ket{0} - e^{-i (\Omega t+\varphi)} \sin \theta \ket{1} \ ,  \\
 \ket{\phi_2} &= \sin \theta \ket{0} e^{-i \varphi} + e^{-i \Omega t} \cos \theta \ket{1}  
\end{split} \label{eq:floquet_state}
\ee
and the corresponding quasienergies are
\be
\epsilon_{1,2} =  \frac{1}{2}(\Delta \pm \Omega_R) \ ,  \label{eq:parameters}
\ee
where we have introduced the quantities $\Delta = \omega - \Omega$, $\tan 2 \theta = 2 g/\Delta$ and $\gap =  \sqrt{ \Delta ^2 +4 g^2}$.

In order to write the GQME \eqref{eq:fsa_gqme}, we need to calculate the matrix elements of the noise operator $S$ in \eqref{eq:S_al,be,k}.
We consider an Ohmic spectral density $J(\omega) = \eta \omega$
(we assume that $\Omega$ is much smaller than the cutoff frequency of the environment).
Notice from \eqref{eq:spectral_density} that for $E>0$ we have $s(E)/s(-E) = e^{-\beta E}$.

\subsection*{Transverse coupling ($S=\sigma_x$)}

For $S=\sigma_x$ only the terms with $k=\pm 1$ in \eqref{eq:S_al,be,k} contribute and we have
\be 
\begin{split}
   S_{11,-1} &=  S_{11,1} = -  S_{22,-1} = -  S_{22,1} = - \frac{1}{2} \sin{2 \theta} \ , \\
   S_{12,-1} &=   S_{21,1}= \cos^2{\theta} \ , \\
   S_{21,-1} &=   S_{12,1}= -\sin{2 \theta} \ .
\end{split}
\ee

From these expression the calculation follows the one leading to \eqref{eq:PofE_app} with the definitions \eqref{eq:a_defs}. 

\subsection*{Longitudinal coupling ($S=\sigma_z$)}

For $S=\sigma_z$ only the terms with $k=0$ contribute and we have
\be
\begin{split}
  S_{11,0} &= - S_{22,0}= \cos{2 \theta} \ ,  \\
  S_{12,0} &=  S_{21,0}= \sin{2 \theta}  \ .
  \end{split}
\ee
In the case, the GQME reads
\be
\begin{split}
  \dot{\rho}_{11}^\nu &=  2 \pi  S_{12,0}^2 \left[-\rho_{11}^\nu s(\gap )+\rho _{22}^\nu  e^{i
   \gap  \nu } s(-\gap )\right] \ ,  \\
   \dot{\rho}_{22}^\nu &=  2 \pi  S_{12,0}^2 \left[-\rho _{22}^\nu  s(-\gap )+\rho _{11}^\nu  e^{-i
   \gap  \nu } s(\gap )\right].
   \label{eq:GME_sigma_z}
\end{split}
\ee
The GQME \eqref{eq:GME_sigma_z} can be solved analytically. 
By using the notation $\Gamma_{\pm} = 2 \pi  S_{12,0}^2 s(\pm \gap )+s(\gap )$ and $\Gamma = \Gamma_{+}+ \Gamma_{-} $, the CF reads
\be 
\begin{split}
 \CF &= \frac{  \Gamma_+}{\Gamma} \rho _{11}(0)  \left(1-e^{- \Gamma t}\right) e^{-i \gap  \nu }  \\ 
 &+ \frac{  \Gamma_-}{\Gamma} \rho _{22}(0)  
   \left(1-e^{- \Gamma t}\right) e^{i \gap  \nu }  \\
  &+ \frac{  \Gamma_+}{\Gamma}  \left[\rho _{22}(0)+\rho _{11}(0) e^{- \Gamma t}\right] \\
  &+\frac{  \Gamma_-}{\Gamma} \left[\rho _{11}(0)+\rho _{22}(0) e^{- \Gamma t}\right].
   \label{eq:G_lambda_sigma_z}
\end{split}
\ee
From this CF we see that the moments have a particular symmetry.
As $\partial^k/ \partial \nu^k \Big(e^{\pm i \gap  \nu } \Big)=  (\pm i\gap)^k e^{\pm i \gap  \nu }$, the $k-$th moment reads
\begin{equation}
  \average{Q^k} \equiv \frac{1}{i^k} \frac{\partial^k \CF}{\partial \nu^k } \Big|_{\nu=0} 
  =  \gap^k \Big[ p_\downarrow(t) + (-1)^k p_\uparrow(t) \Big]
  \label{eq:Q^k}
\end{equation}
where
\be
\begin{split}
 p_\downarrow(t) &= \frac{  \Gamma_-}{\Gamma} \rho_{22}(0)  \left(1-e^{- \Gamma t}\right) \ ,   \\
 p_\uparrow(t) &= \frac{  \Gamma_+}{\Gamma} \rho_{11}(0)  \left(1-e^{- \Gamma t}\right) \ .  
\end{split}
\ee 
The PDF is immediately obtained from the inverse Fourier transform of the CF.
It reads
\begin{equation}
 P(Q) = p_\downarrow(t) \delta(Q-\gap) + p_\uparrow(t) \delta(Q+\gap) + p_0(t) \delta(Q) \ ,
 \label{eq:Pe_sigma_z}
\end{equation}
where $p_\downarrow(t)$ is the probability for the environment to absorb a
photon (the system emits it), $p_\uparrow(t)$ to emit a photon (the
system absorbs it), and $p_0(t) = 1- p_\downarrow(t) - p_\uparrow(t)$ corresponds to no emissions or
absorptions.

\subsection*{Dependence on the initial state}

The moments $ \average{Q^k}$ in \eqref{eq:Q^k} depend on the initial state.
If the system is initially thermalized in the Floquet basis, i.e., $\rho _{22}(0) = e^{-\beta  \gap}\rho _{11}(0)$, 
by using the relation $s(\gap ) = s(-\gap ) e^{-\beta  \gap}$, we verify that  
\begin{equation}
  \average{Q^k} = 
  \left \{ \begin{array}{cc}
 0 &~{\rm for~ } k ~{\rm  odd} \ , \\
  \gap^k  \frac{2 \left(1-e^{-\Gamma t}\right)  e^{\beta  \epsilon}}{(1+e^{\beta  \epsilon})^2}  &~{\rm for~ } k ~{\rm  even} \ .  \\
\end{array}
\right .\
   \label{eq:sigma_z_moments_floquet}
\end{equation}
That means that $\average{Q}=0$ and, in general, $\average{Q^{2k+1}}=0$.
From this result [or directly from \eqref{eq:Q^k}], it is immediate to see that the average heat power $\mean{\dot{Q}} = 0$ at DSS, as discussed in the main text.

If the system is initially in the state $\ket{\psi_0} = \cos \delta \ket{0} - e^{i \gamma} \sin \delta \ket{1}$, the corresponding populations in the Floquet basis are 
\begin{equation}
 \rho_{11}(0) =\frac{1}{2} [1+\cos 2
   \delta  \cos 2 \theta  -\sin 2 \delta  \sin 2 \theta  \cos (\gamma +\varphi )] 
\end{equation}
and $\rho_{22}(0)=1-\rho_{11}(0)$.
For a resonant drive, i.e., $\theta=\pi/4$, $\delta=\pi/4$ and $\gamma=0$, we have $\rho_{11}(0) =1/2(1-\cos \varphi)$.
In the long time limit, we have 
\begin{equation}
 \frac{\average{Q}}{\gap} = \frac{  s(-\gap )  }{s(\gap )+s(-\gap )}  - \frac{1}{2}(1-\cos \varphi) \ .
\end{equation}
In the low temperature regime, i.e., $s(\gap )/s(-\gap ) \ll 1$, we have $\average{Q}/\gap = 1/2(1+\cos \varphi)$, which vanishes for $\varphi=\pi$.
In the high temperature regime, i.e., $s(\gap )/s(-\gap )\approx 1$, we have $\average{Q}/\gap =1/2 \cos \varphi$. which has no definite sign and attains its minimum $\average{Q}/\gap= -1/2$ at $\varphi=\pi$.
A negative $\average{Q}/\gap$ implies a net heat flow from the environment into the system.

\section{GQME in the time-independent case}
\label{app:GQME_time_indep}

For completeness we also present a derivation of the GQME for an undriven
two-level system, starting from \eqref{eq:gqme_schro}.
We take $H=\frac{\omega}{2} \sigma_z$, $\omega>0$, and $S=\sigma_x$. This gives
$S(-\tau)=e^{i \omega \tau} \sigma_- + e^{-i \omega \tau} \sigma_+ $.
Substituting in \eqref{eq:gqme_schro}, we obtain
\begin{equation}
\begin{split}
\frac{\partial}{\partial t} \rho^\nu(t) = & - i \left[ H, \rho^\nu \right]  
	-  \int_0^\infty d\tau \\
& g(\tau) \sigma_x [e^{i \omega \tau} \sigma_- + e^{-i \omega \tau} \sigma_+] \rho^\nu \\ 
- & g(-\nu+\tau) [e^{i \omega \tau} \sigma_- + e^{-i \omega \tau} \sigma_+] \rho^\nu \sigma_x \\
- & g(-\nu-\tau) \sigma_x  \rho^\nu [e^{i \omega \tau} \sigma_- + e^{-i \omega \tau} \sigma_+] \\ 
+ & g(-\tau) \rho^\nu [e^{i \omega \tau} \sigma_- + e^{-i \omega \tau} \sigma_+]  \sigma_x \ . 
\end{split}
\end{equation}
Using \eqref{eq:g_int}, we obtain
\begin{equation}
\begin{split}
\frac{\partial}{\partial t} \rho^\nu(t) = & - i \left[ H, \rho^\nu \right] \\
- & \sigma_x [I_+(0,-\omega) \sigma_- + I_+(0,\omega) \sigma_+] \rho^\nu \\ 
+ & [I_+(-\nu,-\omega) \sigma_- + I_+(-\nu,\omega) \sigma_+] \rho^\nu \sigma_x \\
+ & \sigma_x  \rho^\nu [I_-(-\nu,-\omega) \sigma_- + I_-(-\nu,\omega) \sigma_+] \\ 
- & \rho^\nu [I_-(0,-\omega) \sigma_- + I_-(0,\omega) \sigma_+]  \sigma_x \ .
\end{split}
\end{equation}
We define the rates $d= \pi J( \omega) \left[ n_B( \omega)+1\right]$ and $u= \pi J( \omega) n_B( \omega)$. Then
\begin{equation}
\begin{split}
\frac{\partial}{\partial t} \rho^\nu(t) = & - i \left[ H, \rho^\nu \right] \\
- & \sigma_x [ d  \sigma_- + u \sigma_+] \rho^\nu \\ 
+ & [e^{- i \omega\nu} d \sigma_-
+ e^{ i \omega\nu} u
\sigma_+] \rho^\nu \sigma_x \\
+ & \sigma_x  \rho^\nu [e^{ i \omega \nu} u
\sigma_-
+ e^{-i \omega\nu} d
\sigma_+] \\ 
- & \rho^\nu [u \sigma_- + d \sigma_+]  \sigma_x \ . 
\end{split}
\end{equation}
The time-independent differential equations for the relevant matrix elements read
\begin{equation}
\begin{split}
\dot \rho_{11}^\nu & = -2 u ~\rho_{11}^\nu+2 d ~e^{i  \omega \nu } \rho_{22}^\nu \ , \\
\dot \rho_{22}^\nu & = 2 u~ e^{-i  \omega \nu } \rho_{11}^\nu-2 d~ \rho_{22}^\nu \ .
\end{split} \label{eq:gqme_indep_001}
\end{equation}
The equations \eqref{eq:gqme_indep_001} have the same structure as the \eqref{eq:GME_sigma_z}.
We thus obtain the same CF and PDF, with the substitutions $ \omega \rightarrow \gap$, $d\rightarrow \pi S_{12,0}^2 s(-\gap)$ and $u \rightarrow \pi S_{12,0}^2 s(\gap)$.
	

\end{document}